# Astro2020 Science White Paper

# The Radio Universe at Low Surface Brightness: Feedback & accretion in the circumgalactic medium

**Thematic Area:** Galaxy Evolution

**Principal Authors:**
Name:           Bjorn Emonts
Institution:    National Radio Astronomy Observatory
Email:          bemonts@nrao.edu
Phone:          434-296-0391

Name:           Mark Lacy
Institution:    National Radio Astronomy Observatory
Email:          mlacy@nrao.edu
Phone:          434-244-6803

**Co-authors:**
Kristina Nyland (NRC, resident at NRL), Brian Mason (NRAO), Matthew Lehnert (IAP), Chris Carilli (NRAO), Craig Sarazin (Univ. Virginia), Zheng Cai (Tsinghua Univ./Lick Obs.), Suchetana Chatterjee (Presidency Univ.), Helmut Dannerbauer (IAC), John Gallagher (Univ. Wisconsin-Madison), Kevin Harrington (Univ. Bonn/MPI), Desika Naryanan (Univ. Florida/Univ. Copenhagen), Dominik Riechers (Cornell Univ./MPI), Graca Rocha (JPL)

**Abstract**:
Massive galaxies at high-z are known to co-evolve with their circumgalactic medium (CGM). If we want to truly understand the role of the CGM in the early evolution of galaxies and galaxy-clusters, we need to fully explore the multi-phase nature of the CGM. We present two novel science cases that utilize low-surface-brightness observations in the radio regime to better understand the CGM around distant galaxies. At the lowest temperatures, observations of widespread molecular gas are providing evidence for the cold baryon cycle that grows massive galaxies. At the highest temperatures, observations of the Sunyaev-Zeldovich Effect are starting to reveal the effect of quasar feedback onto the hot gas in the CGM. We discuss the critical role that radio interferometers with compact configurations in the millimeter regime will play over the next decade in understanding the crucial role of the multi-phase CGM in galaxy evolution.

# 1. Background

Massive galaxies and quasars at high redshifts are known to co-evolve with their circumgalactic medium (CGM) through a variety of processes, from stream-fed accretion of gas from the Cosmic Web to galaxy mergers, and to feedback from starburst and Active Galactic Nuclear (AGN) activity (e.g., Narayanan et al. 2015; Angles-Alcazar et al. 2017; Faucher-Giguere et al. 2016). This makes the CGM in the Early Universe a complex environment, where the interaction and mixing of gas from various physical processes is likely very efficient (Cornuault et al. 2017).

Our main window to explore the CGM at high redshifts has been the optical and near-infrared regime, which at z≥2 includes the strong restframe-ultraviolet Lyα emission of warm gas ($T \geq 10^4$ K). Advances in integral-field spectrographs, such as the Keck Cosmic Web Imager (KCWI) and VLT Multi Unit Spectroscopic Explorer (MUSE), are offering exciting opportunities to increase our understanding of the low-surface-brightness universe, by studying the warm circum- and intergalactic medium across hundreds of kpc (e.g., Martin et al. 2015; Swinbank et al. 2015; Borisova et al. 2016; Cai et al. 2018). In addition, optical absorption-line studies revealed that the halos of massive quasars contain cool, metal-rich gas beyond the typical virial radius (~160 kpc; Hennawi et al. 2006). This gas has a high covering fraction of optically thick HI absorbers and a clumpy morphology with a high volume density (Prochaska et al. 2014; Lau et al. 2016). Some of the strongest absorption line systems also show absorption in the Warner and Lyman bands of $H_2$ (e.g., Balashev et al. 2018; Krogager & Noterdaeme 2018).

Despite these exciting advances, a major limitation in our understanding of the multi-phase CGM around distant galaxies and quasars is that we have little information about both the hottest and coldest gas in the CGM. Hot X-ray emitting gas ($T \sim 10^{7-9}$ K) has been observed around distant radio galaxies and clusters (e.g., Carilli et al 2003; Fassbender et al. 2011), but studying accretion and feedback processes in X-rays are challenging, due to the tenuous nature of the hot gas and often the presence of a bright AGN. Contrary, while ALMA has opened up studies of the coldest (~10-100 K) gas phase at (sub)-kpc scales within distant galaxies and AGN (e.g., Calistro Rivera et al. 2018; Spilker et al. 2018), little is known about cold molecular gas on scales of the CGM. Since molecular gas is the raw ingredient for star formation, a directly connection between the CGM and the stellar build-up of high-z galaxies thus remains missing.

# 2. The Radio Universe at low surface brightness

In this white paper, we present two science cases that are opening up interesting windows for exploring the multiphase CGM with radio interferometers over the next decade. Both science cases are explicitly based on exploring the *low-surface-brightness Universe in the radio regime*.

## 2.1 Baryon recycling via the cold molecular medium

A number of studies have started to detect cold molecular gas on scales of tens of kpc in the circumgalactic environments of massive galaxies at z~2-3 (Emonts et al. 2014, 2016; Cicone et al. 2015; Ginolfi et al. 2017; Dannerbauer et al. 2017). The bright carbon monoxide lines, CO(*J*, *J*-1), have long been the preferred tracer for high-z studies of molecular gas. The ground-transition CO(1-0) ($v_{rest}$ = 115.27 GHz) is the most robust CO-tracer of cold molecular gas, in particular for widespread and sub-thermally excited gas in the CGM. The reason is that CO(1-0) has an effective critical density of only several 100 cm$^{-3}$, and the *J* = 1 level is populated down to *T* ~ 10 K. The higher *J*-transitions of CO trace increasingly warmer and denser gas, for example in regions of star formation and AGN activity (see, e.g., Papadopoulos et al. 2004).

An intriguing case for the existence of a cold molecular medium at high-z is the Spiderweb Galaxy (Fig. 1). Sensitive surface-brightness observations of CO(1-0) with the Australia Telescope Compact Array (ATCA) and Karl G. Jansky Very Large Array (VLA) revealed a widespread (~70 kpc) reservoir of cold molecular gas across the CGM, with a mass of $M_{H2} \sim 10^{11}$ ($\alpha_{CO}/4$) $M_\odot$ (with $\alpha_{CO}=M_{H2}/L'_{CO}$). The molecular gas follows diffuse blue light from in-situ star formation in the CGM (Hatch et al. 2008). Since the gas density and star-formation rate fall along the Schmidt-Kennicutt relation, this provided the first direct link between star formation and molecular gas in the CGM at z~2 (Emonts et al. 2016). Complementary ALMA data of [CI] $^3P_1$-$^3P_0$, CO(4-3) and $H_2O$ showed that the CGM has a carbon abundance and excitation properties similar to the interstellar medium (ISM) in starforming galaxies (Emonts et al 2018a) and that the propagating radio jets locally cool gas in the CGM (Gullberg et al. 2016). This suggests that recycling and mixing through metal-enriched outflows or mass transfer among galaxies occurs on a massive scale in the CGM of the Spiderweb.

These results are starting to reveal a picture that the CGM is truly multi-phase, and that the Early Universe may contain much more molecular gas hiding outside galaxies than has thus far been observed. Even at z~5-7, a recent stacking analysis revealed that [CII] emission from mostly cold neutral gas in the halo is commonly extended well beyond the stellar continuum in HST imaging (Fujimoto et al. 2019). However, important questions remain about the nature of the cold molecular CGM, and its role in galaxy evolution. How common is the presence of a cold CGM, in particular among the more general population of high-z galaxies? Is the cold gas diffuse or associated with star-formation? What is its origin (tidal debris, outflows, cooling, mixing, or accretion)? And how can we observe this cold CGM? These are major questions on galaxy evolution that should be addressed in the next decade.

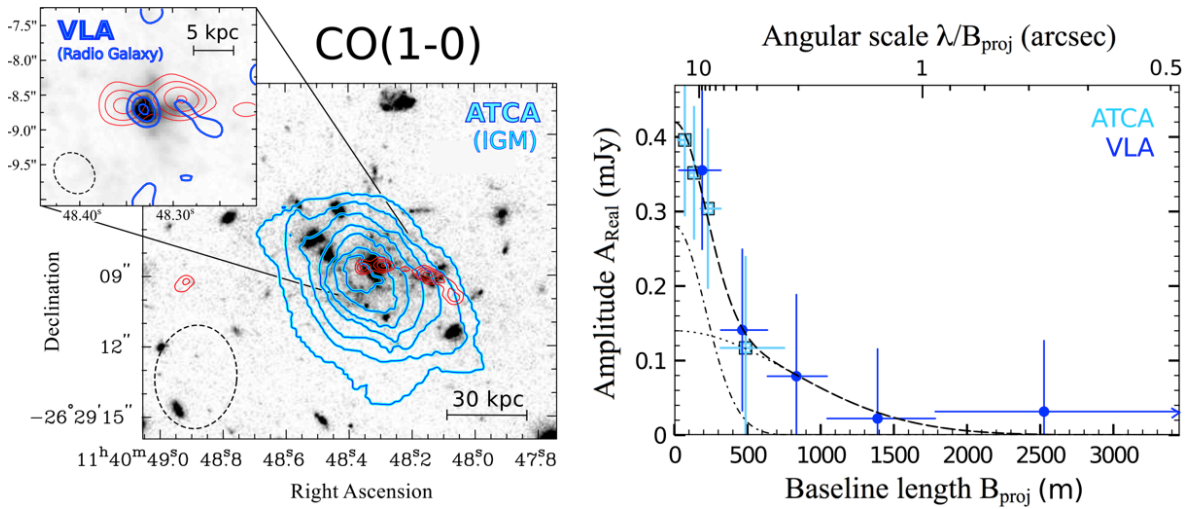

*Figure 1.* CO(1-0) across the CGM of the Spiderweb Galaxy (z=2.2). Left: ATCA total-intensity CO(1-0) contours (blue; 0.020 – 0.128 Jy bm$^{-1}$ × km s$^{-1}$) overlaid onto an HST/ACS image (Miley et al. 2006). The inset shows the total-intensity CO(1-0) contours from the VLA in DnC+CnB-configuration at 3.5, 4.5σ. The untapered VLA data reveal only 1/3rd of the CO imaged on large scales with ATCA. Right: Visibility-amplitudes of the CO(1-0) plotted against the projected baseline length. The dashed line shows a model with superimposed Gaussians distributions of FWHM = 3″ (dash-dotted line) and FWHM = 0.8″ (dotted line). Figure adapted from Emonts et al. (2016).

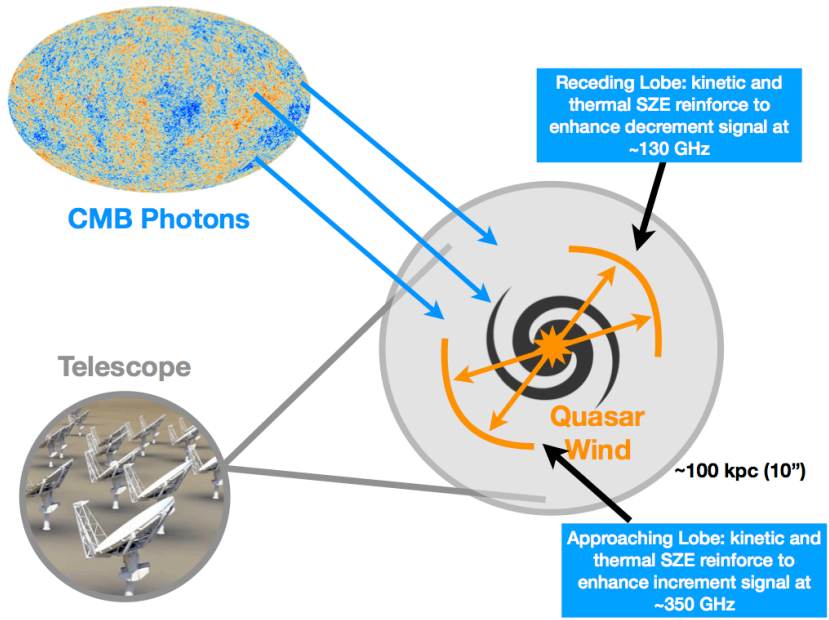

*Figure 2. schematic of the imaging of the SZE from a quasar wind. The combination of the kinetic and thermal SZE components means that winds typically appear asymmetric, with the decrement signal being enhanced at low frequencies in the receding lobe, and the increment signal being enhanced in the approaching lobe at high frequencies. Figure adapted from Lacy et al. (2018).*

## 2.2 AGN Feedback via the Sunyaev-Zeldovich Effect

Winds from AGN and starbursts are one of the major sources of feedback in galaxy evolution. Their interaction with the ISM of the galaxy can inject turbulence, dissociate molecular gas, and/or drive the gas out of the galaxy completely (Silk & Rees 1998; Bower et al. 2006; Hopkins et al. 2006). Thus, quasar winds are likely to be one of the mechanisms by which the CGM is refreshed with material from the host galaxy, and their lifetimes and kinetic luminosities are therefore important to determine. Estimating the energetics in such winds is very difficult, however, as, in most models, the dominant phase in the outflowing gas is hot (~$10^7$ K) gas with low density. Emission from such gas can only be detected in the X-ray. The dependence of X-ray emission on the square of the density, and confusion with point sources such as AGN, make any detection extremely difficult (e.g., Powell et al. 2018). There is another way to detect the host gas in these winds other than through their emission, and that is through the Sunyaev-Zeldovich Effect (SZE). This is the distortion to the spectrum of the Cosmic Microwave Background radiation when it travels through hot gas. Two terms exist, one thermal and one kinetic. The thermal component is proportional to the line of sight electron pressure through the gas, while the kinetic component is proportional to the line of sight bulk velocity of the electrons. These two components provide robust physical measurements of the outflow properties, with the thermal component proportional to the integrated electron temperature and the kinetic component proportional to the line-of-sight bulk velocity of the electrons. The SZE signal is independent of redshift, making it a powerful tool for studying the intergalactic medium (IGM) at high-z.

A wind from a quasar or starburst produces an expanding bubble of hot gas that is highly overpressured compared to the surrounding medium (interstellar through circumgalactic), and moving with respect to the frame of the microwave background (Fig. 2). Thus, as first suggested by Natarajan & Sigurdsson (1999), it should be possible to detect the SZE towards these winds (e.g. Chatterjee & Kosowsky 2007; Rowe & Silk 2011). Statistical studies using stacked data from single dish telescopes have detected significant signals from quasar hosts (Chatterjee et al. 2010; Ruan et al. 2015; Crichton et al. 2016), and direct detections are now just possible with deep ALMA observations (Lacy et al. 2019; Fig. 3). Direct detections are crucial, as observations of a resolved wind via the SZE can constrain the kinetic luminosity of the wind, its age, and the lifetime of the hot bubble that the wind blows in the CGM/IGM. These quantities can be fed back into simulations to assess the overall impact of the winds on galaxy formation and evolution.

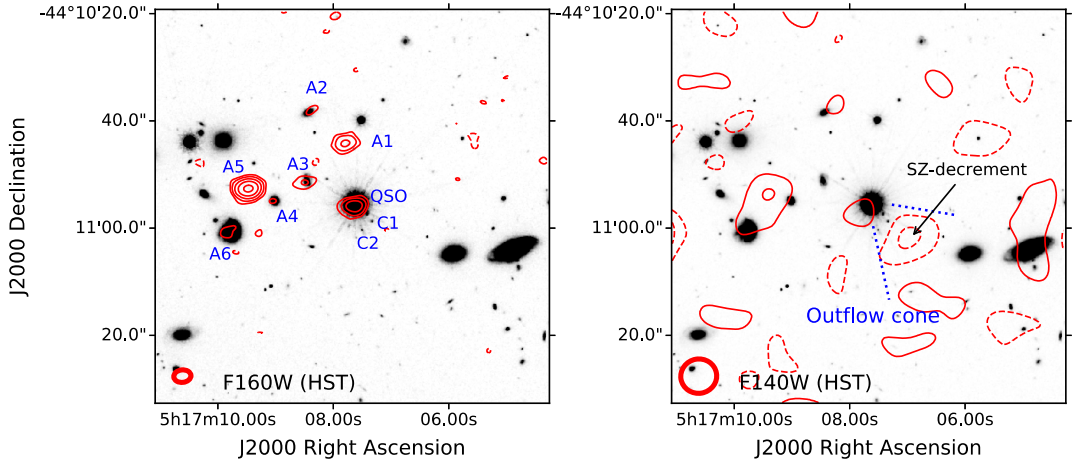

*Figure 3.* *The first detection of the SZE from a quasar wind. The left-hand panel shows the 140 GHz ALMA image at its natural resolution (~2"), the right-hand image has had the point sources removed and been tapered (smoothed) to a resolution of 6.5" to show the SZE detection. The one-sided decrement is consistent with a wind with a kinetic luminosity of about $10^{-4}$ times the bolometric luminosity of the quasar. Figure from Lacy et al. (2019).*

## 3. The *core* aspect of the next-generation radio telescopes

A current challenge for the science in Sect. 2 is that our view of the Radio Universe is biased by the so-called "short-spacing" problem. Because an interferometer samples the sky at discrete intervals that are set by the length of the antenna separations ("baselines"), large-scale features are filtered out and rendered undetectable by widely separated antennas. This is particularly relevant in the mm-regime. For example, baselines shorter than 1 km are needed to recover emission on scales larger than ~2″ (~0.4″) at 35 (140) GHz, which corresponds to ~15 (~4) kpc at z~2.[1] The Radio Universe at low-surface-brightness can only be imaged with array configurations that are densely packed with enough antennas on short baselines to provide the required brightness sensitivity to detect the faint emission from the high-z CGM (see Fig. 1).

To detect the cold molecular CGM in CO(1-0) around the general population of high-z galaxies, surface brightness sensitivities of several milli-Kelvin are needed across a ~100 km s$^{-1}$ line width. Fig. 4 shows a simulation of CO(1-0) from a galaxy at z=2, based on a next-generation Very Large Array (ngVLA). Operating at frequencies between ALMA and the Square Kilometre Array (SKA), i.e. 1.2 – 116 GHz (Murphy et al. 2019), an ngVLA could trace the ground-transition of CO throughout the history of the Universe (Decarli et al. 2018). In the current reference design, the inner 1-km core would be densely packed with 94 antennas of 18m diameter (Carilli et al 2016), providing an order of magnitude more baselines than the current VLA in D-configuration. An additional Short Baseline Array of nineteen 6m antennas and total power capability would further increase brightness sensitivities (Mason et al. 2018). The ngVLA core will be a prime instrument for detecting low-surface-brightness CO in the Early Universe down to a few mK (Emonts et al. 2018b). This can be complemented with ALMA observations of high-*J* CO and other atomic and molecular species (e.g., [CI], [CII], $H_2O$, etc.). ALMA's most compact 12m configuration has maximum baselines of only ~160m, which at these higher frequencies nicely matches, in kλ, the core configuration of an ngVLA. Particularly powerful in preparation for an ngVLA will be to use ALMA's very compact configurations together with the upcoming Band 1 (35-50 GHz), which will allow imaging of CO(1-0) at large angular scales up to z≤2.3. Making a census of the molecular Universe in the radio regime, from the smallest to the largest scales, will also complement high-z observations of warm $H_2$ with JWST.

---

[1] This problem is somewhat alleviated for line emission that is kinematically resolved.

For the case of imaging quasar and starburst winds with the SZE, surface brightness sensitivities of order a few micro-Kelvin in continuum at both ~1mm and 2mm or 3mm are needed on angular scales of ~5″ for a typical quasar or luminous starburst with $L_{bol}$~$10^{12} L_\odot$. Such sensitivities are not easily obtainable with current facilities. Even the SZE from wind bubbles excavated by the most luminous quasars are typically only a few tens of micro-K, and are a challenge to detect even in long duration ALMA observations (Lacy et al. 2019). Increased collecting area on short (~10-100 kλ) baselines at 100-350 GHz, as well as bandwidth to detect the broadband continuum signal, can help here. For example, an enhanced ngVLA core design that allows tapering to 5″ resolution at 100 GHz with only a factor ~2 loss in sensitivity per beam would provide excellent thermal SZE imaging. An ALMA upgrade that includes 2-3× more collecting area and bandwidth would deliver the needed 1mm sensitivity to greatly increase the number of wind bubbles that can be imaged using the SZE. Imaging above 220 GHz is needed to separate the thermal and kinetic SZE, providing critical physical information about the outflows.

The trade-off between an array with a large numbers of relatively small dishes and a large single dish telescope should also be considered. The Green Bank Telescope (GBT) may detect CO(1-0) across the largest scales of the IGM (Harrington et al. 2019) and its 10″ resolution at 100 GHz is just sufficient to resolve a large SZE wind bubble. A single dish will not provide the detailed structural information, and will need a stable bandpass response for high-z spectral-line work, but it can provide surveys for follow-up with interferometers. For studying the SZE signal in emission, a single dish may be more useful at high frequencies. The AtLAST concept, for example (http://atlast.pbworks.com), could provide ~5″ resolution at 350 GHz.

**Through low-surface-brightness radio techniques, the next decade will open these exiting new windows to studying the role of the multi-phase CGM in the early evolution of galaxies.**

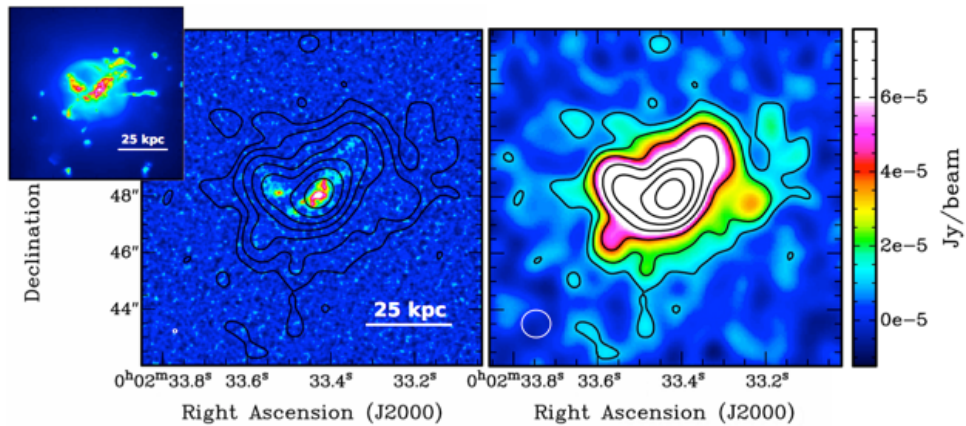

*Figure 4.* CASA simulation of CO(1-0) in a merging system (top-left; Narayanan et al. 2015), using the ngVLA reference design of Carilli (2018) and z = 2, $t_{exp}$ = 48h, 70 km s$^{-1}$ channels, and $I_{CO(1-0)}$ = 0.21 Jy beam km s$^{-1}$. A noise of σ = 2.3 μJy bm$^{-1}$ has been added to the visibilities. Left: Simulated CO(1-0) image at 0.126″ (~1 kpc) resolution, effectively using baselines ≤30 km. The 3σ brightness limit is 0.5 K. Right: Simulated CO(1-0) image at 1″ resolution, relevant to just the km-scale core of the array. The 3σ surface-brightness limit is 0.9 mK and ~60% more CO(1-0) flux is recovered. Figure from Emonts et al (2018b). Credit: Carilli & Erickson 2018 (original).